\documentclass[journal, final]{IEEEtran}
\usepackage{subfigure}
\usepackage[dvips]{graphicx}
\usepackage{amsmath}
\usepackage{cite}
\usepackage{color}
\usepackage{gensymb}
\usepackage{graphicx}

\begin{document}

\title{Continuously Tunable Dual-mode Bandstop Filter}

\author{Amir~Ebrahimi,~\IEEEmembership{Member,~IEEE,}
        Thomas~Baum,~\IEEEmembership{Member,~IEEE,}\\
        James~Scott,~\IEEEmembership{Member,~IEEE,}
        and Kamran~Ghorbani,~\IEEEmembership{Senior~Member,~IEEE}
\thanks{The authors are with the School of Engineering, Royal Melbourne Institute of Technology (RMIT), Melbourne, VIC 3001, Australia (e-mail: amir.ebrahimi@rmit.edu.au).}} %

\maketitle

\begin{abstract}
A varactor-based tunable bandstop filter has been proposed in this article. The proposed filter is based on a dual-mode circuit developed by introducing inductive and capacitive couplings into a notch filter. The frequency tunability is achieved by using varactor diodes instead of the lumped capacitors in the circuit. Next, the equivalent circuit model has been implemented in planar microstrip technology using thin inductive traces and varactor diodes. The fabricated filter prototype shows a continuous center frequency tuning range of $0.66-0.99$~GHz with a compact size of $0.12\lambda_g \times 0.16\lambda_g$, where $\lambda_g$ is the guided wavelength at the middle frequency of the tuning range.  
 
\end{abstract}
\IEEEpeerreviewmaketitle
\begin{IEEEkeywords}
Bandstop filter, dual-mode filter, second-order filter, tunable filter.
\end{IEEEkeywords}
\vspace{-6mm}
\section{Introduction}
\IEEEPARstart{N}{owadays}, with the advent of wireless standards using several frequency bands, there is an essential requirement in the design of communication systems with compatibility of tuning or reconfiguring to new frequency bands \cite{Zhang2012}. Bandstop or notch filters are useful blocks in many of these systems for rejecting unwanted bands. Thus, there is a particular requirement in designing tunable bandstop filters for such systems. Variety of configurations and methods have been developed for tunable bandstop filters \cite{Zhang2012,Ou2011,Ko2015,Cho2015,Naglich2010,Lee2013,Reines2010,Chun2009}. The filter in \cite{Zhang2012} incorporates $\lambda/2$ coupled line resonators yielding a large circuit size. In \cite{Ou2011,Ko2015}, lumped-element tunable bandstop filters are designed for cognitive radios. Nevertheless, the quality factor of lumped-elements are limited for higher frequency implementations. Furthermore, three-layer PCB is required for realizing the inductive couplings in \cite{Ko2015} and precise position adjustment should be performed to design the required coupling between the inductors. In \cite{Cho2015}, a high rejection tunable stopband filter is designed using multi-layer circuit board. High quality factor is achieved in \cite{Naglich2010,Lee2013} by using microstrip-line-coupled cavity resonators tuned by piezo actuators. However, they require complicated multilayer fabrication. Micro-electromechanical (MEMS) varactors are used in \cite{Reines2010} for tuning distributed coupled-line bandstop filters. Nevertheless, the coupled line configurations are not compatible with small size devices at lower frequencies. A tunable bandstop filter is designed by tunable defected ground (DG) resonators loaded with BST varactors in \cite{Chun2009}. However, this filter suffers from a low tuning range and high return loss. 

This paper presents a dual-mode varactor tunable two-pole bandstop filter. The dual-mode circuit is designed by introducing inductive and capacitive couplings into the in/output resonators of a notch filter. Thin inductive traces are used for implementation of inductors. In contrast to the previous designs, the inductive coupling is realized by a third inductor embedded in the ground plane of the circuit. This alleviates the need for complicated three-layer fabrication, offers more flexibility in designing the coupling coefficient, and saves the top layer circuit area leading a more compact size. The filter design procedure and the operation principle are explained in the next sections.

\begin{figure}[!t]
\centering
\includegraphics[width=3.5in]{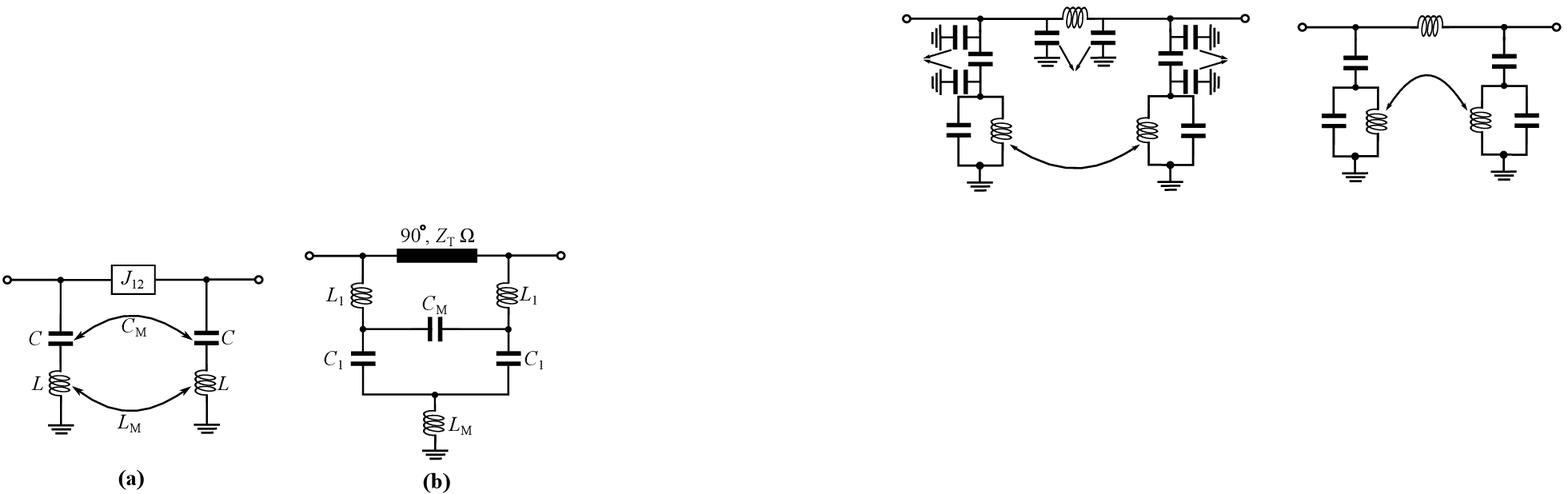}
\caption{Circuit model of the filters. (a) Introducing inductive and capacitive couplings to the notch filter. (b) The proposed bandstop filter after replacing the inductive and capacitive couplings with their equivalent circuit models. }
\label{Fig1}
\end{figure}

\setlength{\textfloatsep}{2pt}

\vspace{-5mm}

\section{Filter Design}
\subsection{Synthesis of the Filter}
\label{sec:Basics}

The proposed filter is designed by applying inductive and capacitive couplings between the series $LC$ resonators of the degenerative-pole notch filter as indicated in Fig.~\ref{Fig1}(a)\cite{Ou2011}. As explained in \cite{Ebrahimi2016}, introducing the couplings separates the two poles of the filter resulting in a dual-mode bandstop filter \cite{Ebrahimi2016,Ebrahimi2014a}. The $J_\mathrm{12}$ inverter can be realized with a $90^\circ$ transmission line. By replacing the inductive and capacitive couplings with their $T$ and $\pi$ equivalent circuits respectively, the circuit model takes the form shown in Fig.~\ref{Fig1}(b). The $L$ and $C$ values in Fig.~\ref{Fig1} can be synthesized as \cite{Zverev1967}  
\setlength{\belowdisplayskip}{0pt} \setlength{\belowdisplayshortskip}{0pt}
\setlength{\abovedisplayskip}{0pt} \setlength{\abovedisplayshortskip}{0pt}
\begin{equation}\label{Eq1}
Z_T=\dfrac{Z_0}{\sqrt{g_0g_3}},
\end{equation}

\begin{equation}\label{Eq2}
L=\dfrac{Z_0}{\Delta \omega_0 \sqrt{g_1g_2}},
\end{equation}

\begin{equation}\label{Eq3}
C=\dfrac{1}{\omega_0^{2} L},
\end{equation}
where $g_0-g_3$ are the element values of the lowpass filter prototype, $\omega_0$ is the central frequency of the filter, $Z_0$ is the termination impedance, and $\Delta$ is the fractional bandwidth. Next, using the standard coupled-resonators filter theory, the element values of the filter in Fig.~\ref{Fig1}(b) are calculated as \cite{Zverev1967}

\begin{equation}\label{Eq4}
L_\mathrm{M}=\Delta k L,
\end{equation}

\begin{equation}\label{Eq5}
L_\mathrm{1}=(1-\Delta k)L,
\end{equation}

\begin{equation}\label{Eq6}
C_\mathrm{M}=\Delta k C,
\end{equation}

\begin{equation}\label{Eq7}
C_\mathrm{1}=(1-\Delta k)C,
\end{equation}

\begin{figure}[!t]
\centering
\includegraphics[width=1.6in]{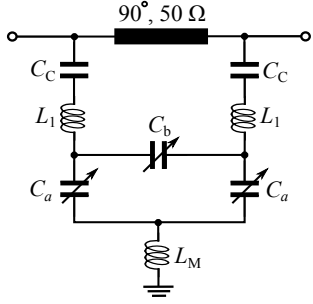}
\caption{Modified circuit model to acheive practical values for the varactors. }
\label{Fig2}
\end{figure}

\setlength{\textfloatsep}{2pt}

where $k$ is the normalized coupling coefficient between the input and output resonators.

\vspace{-4mm}
\subsection{Tunable Filter}
A tunable bandstop filter is designed by replacing the $C_1$ and $C_M$ capacitors with varactor diodes, where the $C_1$, mainly controls the center frequency and $C_M$ is used to control the rejection level \cite{Reines2010}. However, the circuit in Fig.~\ref{Fig1}(b) results in very small values of $C_M$ that might not be achievable using commercially available varactors. To address this challenge, the circuit in Fig.~\ref{Fig2}(a) can be used. In this circuit, series capacitors $C_C$ are introduced to acheive more practical values of the capacitors. By choosing a practically available value for $C_C$, and equating the $T$ counterpart models of the capacitors in Fig.~\ref{Fig1}(b) and Fig.~\ref{Fig2}(a), the $C_a$ and $C_b$ can be obtained as

\begin{equation}\label{Eq8}
C_a=\dfrac{C_k C_j}{2C_k+C_j},
\end{equation}

\begin{equation}\label{Eq9}
C_b=\dfrac{C_k^2}{2C_k+C_j},
\end{equation}
where $C_k$ and $C_j$ are intermediate variables given by 

\begin{equation}\label{Eq10}
C_k=\dfrac{(1+\Delta k)C C_C}{C_C-C(1+\Delta k)},
\end{equation}

\begin{equation}\label{Eq11}
C_j=\dfrac{1-(\Delta k)^2}{\Delta k}C.
\end{equation}
The $C$ value is obtained from (\ref{Eq3}) and the $L_1$ and $L_M$ inductances are the same in Fig.~\ref{Fig2}(a) and Fig.~\ref{Fig1}(b).

\vspace{-4 mm}
\subsection{Planar Implementation of the Tunable Filter}
\begin{figure}[!t]
\centering
\includegraphics[width=2.7in]{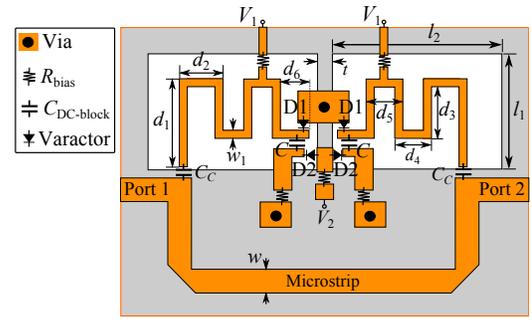}
\caption{Plannar implementation of the tunable filter in microstrip technology. The top microstrip metallization is shown in orange, whereas the ground metallization is gray. The dimensions are: $d_1=11.7$~mm, $w_1~=~0.8$~mm, $d_2=d_4=d_5=4.2$~mm, $d_3=7.1$~mm, $d_6=4.2$~mm, $w=3.5$~mm, $l_1=14.4$~mm, $l_2=20$~mm, $t=0.8$~mm, $l_3=12.5$~mm, $l_4=38.3$~mm.}
\label{Fig3}
\end{figure}

The filter in Fig. 2 can be implemented using lumped components. But, the achievable rejection of the filter would be limited due to the low quality factors of commercial lumped components. Thus, a realization using microstrip is preferable due to lower loss at higher frequencies \cite{Ebrahimi2016}. An implementation in microstrip is shown in Fig.~\ref{Fig3}, where the $L_1$ inductors are implemented with two thin meandered traces with $w_1$ width. The ground metallization is removed beneath these two traces to enhance the $L_1$ inductances.  Also, $L_M$ is realized with the thin metallic trace of $t$ width in the ground plane.  The D1 varactors are considered to implement $C_a$ variable capacitors, and the pairs of D2 series varactors act as the variable $C_b$ capacitors. A via hole connects D1 varactors to $L_M$ in the ground plane. Finally, the microstrip section between the two $C_C$ capacitors is designed to act as a $\lambda/4$ impedance inverter.    

\begin{figure}[!t]
\centering
\includegraphics[width=3.2in]{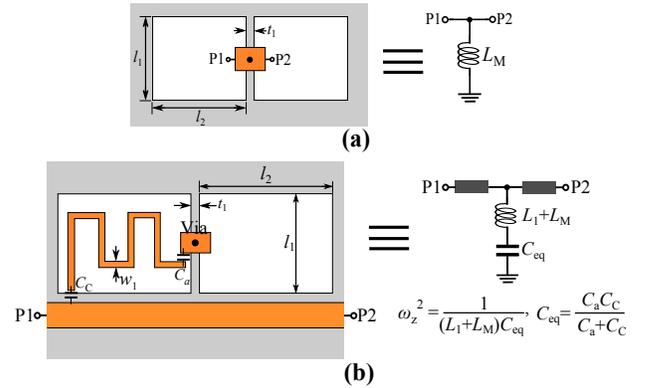}
\caption{(a) Inuctive trace in ground plane realizing $L_M$ and it's circuit model. (b) Circuit for optimizing the dimensions of the $L_1$ trace.}
\label{Fig4}
\end{figure}

\setlength{\textfloatsep}{0.5pt}

\begin{figure}[!t]
\centering
\includegraphics[width=3.5in]{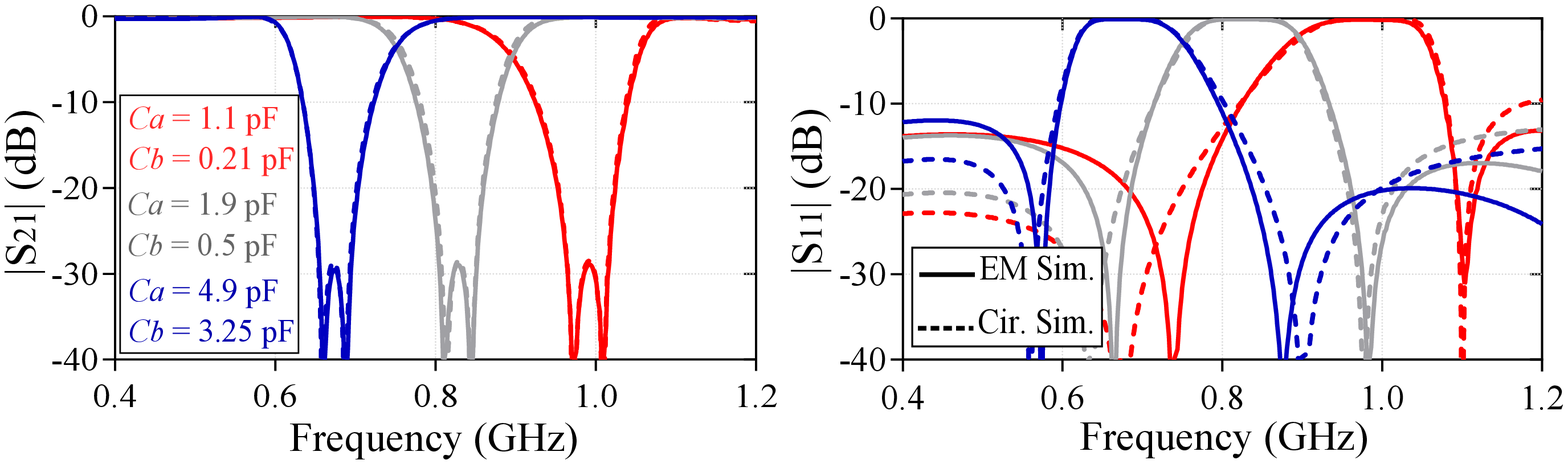}
\caption{Comparison between the EM and circuit model simulations of the designed tunable bandstop filter for three different values of $C_a$, and $C_b$.}
\label{Fig5}
\end{figure}

The design starts by obtaining the element values of the circuit in Fig.~\ref{Fig2} using (\ref{Eq1})-(\ref{Eq11}). The filter should be designed for the center frequency of the tuning band. Then, the tuning is achieved by using varactors instead of the lumped capacitors. The initial dimensions of the metallic trace implementing $L_M$ are obtained using the closed-form equations for microstrip inductors in \cite{Hong2004}. Then, the dimensions are optimized by simulating the structure in Fig.~\ref{Fig4}(a) using \textit{ADS Momentum}, where the $L_M$ is found from the susceptance slope. The next step is to find the dimensions of the thin trace implementing $L_1$. Likewise, equations in \cite{Hong2004} are used as initial values. Then, the dimensions are optimized by simulating the structure in Fig.~\ref{Fig4}(b) and finding the transmission zero frequency $\omega_z$, where $L_1$ is found by knowing $L_M$, $C_a$ and $C_C$. For validation, a Butterworth bandstop filter is designed with a center frequency of $0.83$~GHz and $\Delta=18$\%. The equivalent circuit model parameters by considering $C_C=2.2$~pF are $L_1=L_2=27$~nH, $L_M=3.1$~nH with $C_a=1.9$~pF, and $C_b=0.5$~pF. The layout dimensions are then optimized by curve fitting of the EM and circuit simulations in the \textit{Keysight ADS} software by considering a $Rogers~RO4350$ substrate with $\epsilon_r=3.48$ and $1.524$~mm thickness. A comparison between the EM and circuit model simulations provided in Fig.~\ref{Fig5} shows a good agreement validating the design procedure. The comparison is performed for the center frequencies of $0.66$~GHz and $0.99$~GHz as well, with the corresponding $C_a$ and $C_b$ values given in the inset of Fig.~\ref{Fig4}(a), where all the other circuit parameters are kept unchanged. These results verify the tunability of the designed filter by varying the $C_a$, and $C_b$ capacitance values. 

\begin{figure}[!t]
\centering
\includegraphics[width=3in]{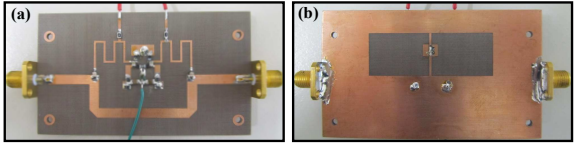}
\caption{The fabricated filter prototype. (a) Front view, and (b) Back view.}
\label{Fig6}
\end{figure}

\begin{figure}[!t]
\centering
\includegraphics[width=2.2in]{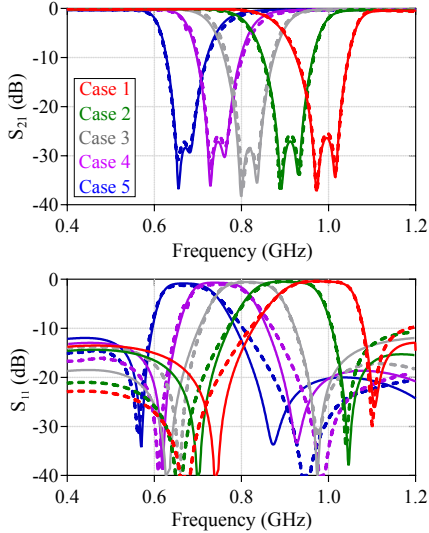}
\caption{Comparison between the simulated (solid line) and measured (dashed line) S-parameters of the filter. The ($V_1,V_2$) values for different cases are: ($15~\mathrm{V},35~\mathrm{V}$) for Case~1, ($8.5~\mathrm{V},30~\mathrm{V}$) for Case~2, ($6~\mathrm{V},11~\mathrm{V}$) for Case~3, ($4.2~\mathrm{V},5~\mathrm{V}$) for Case~4, ($1.7~\mathrm{V},0.2~\mathrm{V}$) for Case~5.  }
\label{Fig7}
\end{figure}

\vspace{-3mm}
\section{Experimental Validation and Results}
\label{Exper}
A prototype of the proposed filter is fabricated and tested for validation. Fig.~\ref{Fig6} shows the front and back views of the fabricated filter. The Infineon BB837 surface-mounted diodes are used as varactors. The biasing networks are made of $R=10$~k$\Omega$ resistors, and $C_\textrm{DC-block}=100$~pF capacitors. Fig.~\ref{Fig7} shows the EM simulated and measured S-parameters of the filter, where the series parasitic resistance of the varactor from the maufacturer's SPICE model is considered in simulations \cite{(2017)}. The filter shows a continuous tuning of the center frequency from $0.66$~GHz to $0.99$~GHz with an almost constant $-3$~dB stopband fractional bandwidth of $18$\%. The measured stopband return loss (SB RL) of the filter is less than $0.8$~dB, whereas the passband insersion loss (PB IL) is less than $0.55$~dB within the tuning range. A relatively low SB RL and PB IL are attributed to the larger bandwidth of the designed filter with respect to the previous ones, using thin meandered microstrip inductors instead of the lumped ones, and using $\lambda/4$ microstrip stub as an impedance inverter instead of lumped inductor. A comparison between the performance of the designed filter and state-of-art designs in Table~\ref{table1} reveals that the designed filter is more compact compared to the topologies implemented with microstrip inductors or distributed resonators. Moreover, the table shows a competitive performance of the proposed filter in terms of the tuning range, and rejection.

\begin{table}
  \renewcommand{\arraystretch}{1.2}
  \caption{Comparison of Various Tunable Bandstop Filters} 
  \label{table1}
  \centering
  \begin{tabular}{c@{\hskip 0.01cm}| c@{\hskip 0.01cm} |c@{\hskip 0.01cm} |c@{\hskip 0.01cm}| c@{\hskip 0.01cm}| c@{\hskip 0.01cm}| c@{\hskip 0.01cm}| c@{\hskip 0.01cm} c@{\hskip 0.01cm}}
  \hline
  \hspace{-0.3cm} Ref. & \begin{tabular}[x]{@{}c@{}}Tun. \\ mech. \end{tabular}& \begin{tabular}[x]{@{}c@{}}\hspace{-0.2cm}Rej. \\\hspace{-0.2cm} (dB) \end{tabular}   & \begin{tabular}[x]{@{}c@{}}Tun. \\ Range (GHz) \end{tabular} &
   \begin{tabular}[x]{@{}c@{}}$\Delta$ \\ (\%) \end{tabular} & \begin{tabular}[x]{@{}c@{}}Size \\ ($\lambda_g^2$) \end{tabular}  & \begin{tabular}[x]{@{}c@{}}SB \\ RL (dB) \end{tabular} & \begin{tabular}[x]{@{}c@{}}PB \\ IL (dB) \end{tabular}\\
  \hline
  \hspace{-0.3cm}\cite{Zhang2012} &\hspace{-0.3cm} Varactor & \hspace{-0.3cm} 50 & \hspace{-0.2cm}1.75--2.2~(11.4\%) & 10 & $0.18$& $2$& $0.65$\\
  \hspace{-0.3cm}\cite{Ou2011} &\hspace{-0.3cm} Varactor &\hspace{-0.3cm} 17&\hspace{-0.2cm} 0.47--0.73~(43\%) & 4.5 &  $0.0004$& $3$& $0.4$\\
  \hspace{-0.3cm}\cite{Ko2015} & \hspace{-0.25cm}Varactor & \hspace{-0.2cm}23 & \hspace{-0.2cm}0.6-0.99~(49\%) & 10 &  $0.005$& N. G. & $1$\\
  \hspace{-0.3cm}\cite{Cho2015} & \hspace{-0.3cm}Varactor & \hspace{-0.2cm}40 & \hspace{-0.2cm}0.76-1.05~(32\%) & 8& $0.034$& $2$ & $1.52$\\
  \hspace{-0.3cm}\cite{Naglich2010}& \hspace{-0.1cm}Piezo Act. & \hspace{-0.2cm}45 &\hspace{-0.2cm} 2.77--3.57~(25.2\%)& 0.9& $0.5$& $2$& $2.4$ \\
\hspace{-0.3cm}\cite{Lee2013} & \hspace{-0.1cm}Piezo Act. & \hspace{-0.2cm}30 & \hspace{-0.2cm} 2.7--3~(10.5\%) & 1.5&  $0.15$& $1.8$& $2$\\
 \hspace{-0.3cm}T. W. &\hspace{-0.3cm} Varactor & \hspace{-0.2cm}27& \hspace{-0.2cm} 0.66-0.99~(40\%) & 18& $0.019$& $0.8$& $0.55$\\
  \hline 
  \end{tabular}
  \end{table}

\vspace{-4mm}
\section{Conclusion}
A second-order tunable bandstop filter has been presented based on a combination of lumped elements and microstrip-based components. An equivalent circuit-based design method has been developed for synthesizing the element values of the filter. The proposed topology offers compact size, high rejection level, and a wide continuous tuning range.



\vspace{-2mm}
\begin{thebibliography}{10}
\providecommand{\url}[1]{#1}
\csname url@samestyle\endcsname
\providecommand{\newblock}{\relax}
\providecommand{\bibinfo}[2]{#2}
\providecommand{\BIBentrySTDinterwordspacing}{\spaceskip=0pt\relax}
\providecommand{\BIBentryALTinterwordstretchfactor}{4}
\providecommand{\BIBentryALTinterwordspacing}{\spaceskip=\fontdimen2\font plus
\BIBentryALTinterwordstretchfactor\fontdimen3\font minus
  \fontdimen4\font\relax}
\providecommand{\BIBforeignlanguage}[2]{{%
\expandafter\ifx\csname l@#1\endcsname\relax
\typeout{** WARNING: IEEEtran.bst: No hyphenation pattern has been}%
\typeout{** loaded for the language `#1'. Using the pattern for}%
\typeout{** the default language instead.}%
\else
\language=\csname l@#1\endcsname
\fi
#2}}
\providecommand{\BIBdecl}{\relax}
\BIBdecl

\bibitem{Zhang2012}
X.~Y. Zhang, C.~H. Chan, Q.~Xue, and B.-J. Hu, ``{RF} tunable bandstop filters
  with constant bandwidth based on a doublet configuration,'' \emph{IEEE Trans.
  Indus. Electr.}, vol.~59, no.~2, pp. 1257--1265, 2012.

\bibitem{Ou2011}
Y.-C. Ou and G.~Rebeiz, ``Lumped-element fully tunable bandstop filters for
  cognitive radio applications,'' \emph{IEEE Trans. Microw. Theory Techn.},
  vol.~59, no.~10, pp. 2461--2468, Oct 2011.

\bibitem{Ko2015}
C.-H. Ko, A.~Tran, and G.~Rebeiz, ``Tunable 500-1200-{MH}z dual-band and wide
  bandwidth notch filters using {RF} transformers,'' \emph{IEEE Trans. Microw.
  Theory Techn.}, vol.~63, no.~6, pp. 1854--1862, June 2015.

\bibitem{Cho2015}
Y.~H. Cho and G.~M. Rebeiz, ``Tunable 4-pole dual-notch filters for cognitive
  radios and carrier aggregation systems,'' \emph{IEEE Trans. on Microw. Theory
  Tech.}, vol.~63, no.~4, pp. 1308--1314, April 2015.

\bibitem{Naglich2010}
E.~J. Naglich, J.~Lee, D.~Peroulis, and W.~J. Chappell, ``A tunable
  bandpass-to-bandstop reconfigurable filter with independent bandwidths and
  tunable response shape,'' \emph{IEEE Trans. Microw. Theory Techn.}, vol.~58,
  no.~12, pp. 3770--3779, 2010.

\bibitem{Lee2013}
J.~Lee, E.~J. Naglich, H.~H. Sigmarsson, D.~Peroulis, and W.~J. Chappell, ``New
  bandstop filter circuit topology and its application to design of a
  bandstop-to-bandpass switchable filter,'' \emph{IEEE Trans. on Microw. Theory
  Techn.}, vol.~61, no.~3, pp. 1114--1123, 2013.

\bibitem{Reines2010}
I.~Reines, S.-J. Park, and G.~M. Rebeiz, ``Compact low-loss tunable $ {X}
  $-band bandstop filter with miniature {RF-MEMS} switches,'' \emph{IEEE Trans.
  Microw. Theory Techn.}, vol.~58, no.~7, pp. 1887--1895, 2010.

\bibitem{Chun2009}
Y.-H. Chun, J.-S. Hong, P.~Bao, T.~Jackson, and M.~Lancaster, ``Tunable slotted
  ground structured bandstop filter with bst varactors,'' \emph{IET Microw.
  Antennas \& Propag.}, vol.~3, no.~5, pp. 870--876, 2009.

\bibitem{Ebrahimi2016}
A.~Ebrahimi, W.~Withayachumnankul, S.~F. Al-Sarawi, and D.~Abbott, ``Compact
  second-order bandstop filter based on dual-mode complementary split-ring
  resonator,'' \emph{IEEE Microw. Wireless Compon. Lett.}, vol.~26, no.~8, pp.
  571--573, Aug 2016.

\bibitem{Ebrahimi2014a}
A.~Ebrahimi, W.~Withayachumnankul, S.~F. Al-Sarawi, and D.~Abbott, ``Dual-mode 
  behavior of the complementary electric-{LC} resonators
  loaded on transmission line: Analysis and applications,'' \emph{J. Appl.
  Phys.}, vol. 116, no.~8, 2014, art. no. 083705.

\bibitem{Zverev1967}
A.~I. Zverev, \emph{Handbook of Filter Synthesis}.\hskip 1em plus 0.5em minus
  0.4em\relax Wiley, 1967.

\bibitem{Hong2004}
J.-S.~G. Hong and M.~J. Lancaster, \emph{Microstrip filters for RF/microwave
  applications}.\hskip 1em plus 0.5em minus 0.4em\relax John Wiley \& Sons,
  2004, vol. 167.

\bibitem{(2017)}
\BIBentryALTinterwordspacing
\emph{Infineon}, (2017). [Online]. Available: \url{https://www.infineon.com}
\BIBentrySTDinterwordspacing

\end{thebibliography}
\end{document}